%% file: publication.tex
\documentclass[letterpaper,aps,prl,twocolumn,showpacs,groupedaddress]{revtex4} 

\usepackage{graphicx}  
\usepackage{dcolumn}   
\usepackage{bm}        
\usepackage{amssymb}   

\newcommand{\pp}  {\mbox{$p\bar{p}$}}
\newcommand{\roots}  {\mbox{$\sqrt{s}=1.96$~TeV}}
\newcommand{\zmumu}  {\mbox{$Z/\gamma^*\rightarrow\mu^+ \mu^-$}}
\newcommand{\ztau}  {\mbox{$Z/\gamma^*\rightarrow\tau^+ \tau^-$}}

\newcommand{\Z}  {\mbox{$Z/\gamma^*$}}

\newcommand{\pt}  {\mbox{$p_{T}$}}
\newcommand{\ptz}  {\mbox{$p_{T}^{Z}$}}
\newcommand{\ptj}  {\mbox{$p_{T}^{\mathrm{jet}}$}}
\newcommand{\ptJ}  {\mbox{$p_{T}^{\mathrm{JET}}$}}
\newcommand{\rap}  {\mbox{$y$}}
\newcommand{\rapz}  {\mbox{$y^Z$}}
\newcommand{\rapj}  {\mbox{$y^{\mathrm{jet}}$}}
\newcommand{\rapJ}  {\mbox{$y^{\mathrm{JET}}$}}
\newcommand{\pythia}  {{\sc pythia}}
\newcommand{\alpgen}  {{\sc alpgen}}
\newcommand{\sherpa}  {{\sc sherpa}}
\newcommand{\mcfm}  {{\sc mcfm}}
\newcommand{\geant}  {{\sc geant}}
\newcommand{\run}  {Run II}

\begin{document}

\hspace{5.2in} \mbox{FERMILAB-PUB-08-293-E}
\title{Measurement of differential {\boldmath $Z/\gamma^*$}+jet+{\boldmath $X$}\ cross sections in {\boldmath $p\bar{p}$}~collisions at {\boldmath $\sqrt{s}=1.96$}~TeV}
\input list_of_authors_r2.tex 
\date{August 8, 2008}

\begin{abstract}
We present new measurements of differential cross sections for \Z($\rightarrow\mu\mu$)+jet+$X$\ production in a  1~fb$^{-1}$\ data sample collected with the D0 detector in \pp\ collisions at \roots.  
Results include the first measurements differential in the \Z\ transverse momentum and rapidity, as well as new measurements differential in the leading jet transverse momentum and rapidity.
Next-to-leading order perturbative QCD predictions are compared to the measurements, and reasonable agreement is observed, except in the region of low \Z\ transverse momentum. 
Predictions from two event generators based on matrix elements and parton showers, and one pure parton shower event generator are also compared to the measurements. 
These show significant overall normalization differences to the data and have varied success in describing the shape of the distributions. 
\end{abstract}


\maketitle

\input{introduction.tex}

\input{selection.tex}

\input{unfolding.tex}

\input{results.tex}

\input{summary.tex}

\input acknowledgement_paragraph_r2.tex
\end{document}

%% file: list_of_authors_r2.tex
%
\author{V.M.~Abazov$^{36}$}
\author{B.~Abbott$^{75}$}
\author{M.~Abolins$^{65}$}
\author{B.S.~Acharya$^{29}$}
\author{M.~Adams$^{51}$}
\author{T.~Adams$^{49}$}
\author{E.~Aguilo$^{6}$}
\author{M.~Ahsan$^{59}$}
\author{G.D.~Alexeev$^{36}$}
\author{G.~Alkhazov$^{40}$}
% ...

\affiliation{\vspace{0.1 in}(The D\O\ Collaboration)\vspace{0.1 in}}
% ...
\affiliation{$^{6}$University of Alberta, Edmonton, Alberta, Canada,
                Simon Fraser University, Burnaby, British Columbia, Canada,
                York University, Toronto, Ontario, Canada, and
                McGill University, Montreal, Quebec, Canada}
% ...
\affiliation{$^{29}$Tata Institute of Fundamental Research, Mumbai, India}
% ...
\affiliation{$^{36}$Joint Institute for Nuclear Research, Dubna, Russia}
% ...
\affiliation{$^{40}$Petersburg Nuclear Physics Institute,
                St. Petersburg, Russia}
% ...
\affiliation{$^{49}$Florida State University, Tallahassee, Florida 32306, USA}
% ...
\affiliation{$^{51}$University of Illinois at Chicago,
                Chicago, Illinois 60607, USA}
% ...
\affiliation{$^{59}$Kansas State University, Manhattan, Kansas 66506, USA}
% ...
\affiliation{$^{65}$Michigan State University,
                East Lansing, Michigan 48824, USA}
% ...
\affiliation{$^{75}$University of Oklahoma, Norman, Oklahoma 73019, USA}
% ...

%% file: introduction.tex
The production of $W$\ or $Z$\ bosons in association with jets is an important signal at hadron colliders such as the Fermilab Tevatron Collider and the CERN Large Hadron Collider. 
Leptonic boson decays can be identified with little background, and measurements of the boson and jet kinematics provide good tests of perturbative QCD (pQCD) calculations and modeling.
Such events also form the main background to many processes with much smaller cross sections, including production of the top quark, Higgs boson, and particles expected  in some supersymmetric scenarios.
Accurate theoretical modeling of $W$\ or $Z$\ boson + jet final states is a key element in studying such rare processes; developing and testing these models relies upon input from experimental measurements of boson + jet production. 

Previous boson + jet measurements at the Fermilab Tevatron Collider~\cite{d0_zjets, cdf_zjets} have included comparisons with next-to-leading order (NLO) pQCD predictions from \mcfm~\cite{mcfm} and a tree-level matrix element calculation matched to a parton-shower (ME+PS) Monte Carlo event generator~\cite{madgraph}.
In this Letter we describe new measurements of differential cross sections in \Z($\rightarrow\mu\mu$) + jet + $X$\ production in \pp\ collisions at \roots, with a data sample corresponding to $0.97\pm0.06$~fb$^{-1}$~\cite{d0lumi}\ recorded by the D0 detector between April 2002 and February 2006.
The measurement is carried out in a region of dimuon mass $65<M_{\mu\mu}<115$~GeV in which the inclusive cross section for \Z\ production is approximately equal to that of pure $Z$\ boson production, and the measured distributions are corrected to the  particle level~\cite{particle_level}.
Differential cross sections binned in the \Z\ momentum component perpendicular (transverse) to the beam, \ptz\ ($\mathrm{d}\sigma_{Z+\mathrm{jet}+X}/\rm{d}\ptz$) and rapidity~\cite{rapidity}, \rapz\ ($\mathrm{d}\sigma_{Z+\mathrm{jet}+X}/\rm{d}\rapz$), are presented here for the first time, and binned in the leading (in \pt) jet \pt\ ($\mathrm{d}\sigma_{Z+\mathrm{jet}+X}/\rm{d}\ptj$) extended to lower \ptj, and \rap\ ($\rm{d}\sigma_{Z+jet+X}/\rm{d}\rapj$)  covering a wider \rapj\ range than previous studies.
Comparisons are made using NLO pQCD predictions from \mcfm\ with non-perturbative corrections applied. 
Currently the best tools for generating simulated boson + jets events are tree-level ME+PS calculations. 
Several such calculations are available using different schemes to combine the matrix element and partons shower contributions, and with some differences in the predicted kinematics~\cite{mc_paper}.
Two such ME+PS event generators are compared to the measurements: \alpgen~\cite{alpgen}, with \pythia~\cite{pythia}\ used for the parton-showering; and \sherpa~\cite{sherpa}.
Finally, a pure parton-shower prediction from \pythia\ is also compared.

%% file: selection.tex
The measurements are made with the D0 detector, which is described in detail elsewhere~\cite{d0det}; a brief overview is given here of the most relevant components for this analysis.
The interaction region is surrounded by a magnetic tracking system,
comprising a silicon micro-strip tracker and a fiber tracker, both located within a 2~T superconducting solenoid magnet. 
Three liquid-argon/uranium calorimeters surround the tracking system: a central section covering pseudo-rapidity $|\eta| \leq  1.1$~\cite{theta}, and two end
calorimeters that extend coverage to $1.1 < |\eta| < 4.2$, each
housed in separate cryostats. 
Scintillators between the cryostats sample shower energy for $1.1<|\eta|<1.4$.
Luminosity is measured using plastic scintillator arrays located in front of the end calorimeter cryostats, covering $2.7 < |\eta| < 4.4$.
A muon system surrounds the calorimetry, consisting of three layers of tracking detectors and scintillation trigger counters; these provide muon identification and triggering for $|\eta|<2$.
Sandwiched between the first and second layer are 1.8~T toroidal iron magnets, allowing an independent momentum measurement in the muon system.

Events used in this measurement are selected with a suite of triggers using information from the muon and tracking systems and are required to have two muon candidates reconstructed in those systems.
The primary collision vertex in each event is reconstructed requiring at least three tracks and applying fit quality cuts. 
To reject mis-reconstructed events and cosmic rays, the muons must be consistent with this vertex in directions both transverse and parallel to the beam.
Based on the information from the tracking system, the muons are required to have $\pt>15$~GeV and dimuon mass $65<M_{\mu\mu}<115$~GeV.

Jets are reconstructed from clusters of energy deposited in calorimeter cells using the D0 \run\ midpoint cone algorithm~\cite{d0jets} with a splitting/merging fraction of 0.5 and a cone size of $\Delta{\cal R} = \sqrt{(\Delta\phi)^2 + (\Delta y)^2} = 0.5$, where $\phi$ is the azimuthal angle; jets caused by noise are rejected with quality and shape cuts.
Jets are corrected for the calorimeter response, instrumental out-of-cone showering effects, and additional energy deposits in the calorimeter that arise through detector noise and pile-up from multiple interactions and previous $p\bar{p}$\ bunch crossings.  
These jet energy scale corrections are determined using transverse momentum imbalance in $\gamma$\ + jet events, after the electromagnetic response is calibrated using \Z$\rightarrow ee$\ events.
For clarity in the following discussions, the measured jet transverse momentum and rapidity after these corrections are denoted \ptJ\ and \rapJ, to distinguish them from the particle level quantities \ptj\ and \rapj.

Further selections ensure that the measurement is carried out in regions with high acceptance and well understood detector performance: the muons are required to have $|\eta| < 1.7$, the primary vertex must be within 50~cm of the detector center along the beam direction, and jets are required to have a \ptJ\ $> 20$~GeV and $|\rapJ|<2.8$.
Additionally, events with jets in the \ptJ\ range 15--20~GeV are kept in the sample for studies of the effects of detector resolution.

The main source of background in this analysis is muons from semi-leptonic decays in high energy jets or $W$+jet production.
This is reduced to negligible levels ($<0.5$\%\ of the final sample) by limiting the sum of track momenta and the calorimeter energy allowed around each muon. 
The muons are also required to not overlap with any jet by requiring angular separation $\sqrt{(\Delta\phi(\mu,\mathrm{jet}))^2 + (\Delta\eta(\mu,\mathrm{jet}))^2} > 0.5$. 
Other sources of background (e.g., top quark production, \ztau) are estimated using simulation and found to be negligible ($<0.1$\%).
A total of 59336 \zmumu\ candidate events are selected before jet requirements, of which 9927 contain at least one jet with $\ptJ>20$~GeV passing all selections.

Two simulations of \Z+jets events are used: a sample generated with \pythia\ v6.323, and a sample generated with \alpgen\ v2.05 using \pythia\ for parton showering, both with the \pythia\ underlying event model configured using tune A~\cite{pythia_tune}. 
Both samples are passed through a \geant~\cite{geant} simulation of the detector response.
Real data events from random bunch crossings are overlaid on the simulation to reproduce the effects of multiple $p\bar{p}$\ interactions and detector noise.

The muon trigger efficiency is measured in data and parameterized in terms of variables related to the geometry of the muon and tracking systems. 
The selected events are then corrected on an event-by-event basis for this efficiency, with the average efficiency being ($88.3\pm0.3$)\%, quoting just the statistical uncertainty.
Muon reconstruction, tracking, and isolation efficiencies are measured in the data and in the simulation; scale factors are applied to correct for the differences.
The total systematic uncertainty on the muon trigger and identification efficiency translates into a 5\%\ uncertainty on the measured cross sections, with no significant dependence on the variables studied in this analysis.
Transverse momentum imbalance in  \Z$\rightarrow ee$\ + jet events is studied in data and simulation, and factors applied to the simulation to correct the jet response for any differences.

%% file: unfolding.tex
To extract the differential cross sections, we correct the reconstructed data distributions to particle-level distributions, deriving the corrections from simulation. 
We first select \Z\ plus jet events based on the detector response in simulated events.
These \Z\ and jet variables are compared to quantities independently measured directly from the particles in the simulated events, applying comparable kinematic selections to minimize acceptance effects.
For this particle level selection, the \Z\ mass and kinematics are reconstructed from the generated muons after QED final state radiation (FSR), requiring $|\eta^{\mu}|<1.7$\ and $65<M_{\mu\mu}<115$~GeV.
Jets are reconstructed using the same reconstruction algorithm, now on all final state particles excluding the \Z\ decay products, and selected requiring $|\rapj|<2.8$.
These particle jets are matched to jets reconstructed in the simulation by requiring $\Delta {\cal R} < 0.5$.
We quote results for leading particle jets with \ptj$ > 20$~GeV; however, due to instrumental effects and resolution, measured jets with \ptJ$ > 20$~GeV include significant contributions from particle jets with lower \ptj.
To study this effect, jets at the particle level are reconstructed to very low \pt\ (3~GeV).

\begin{figure}[!h]\center
\includegraphics[width=85mm]{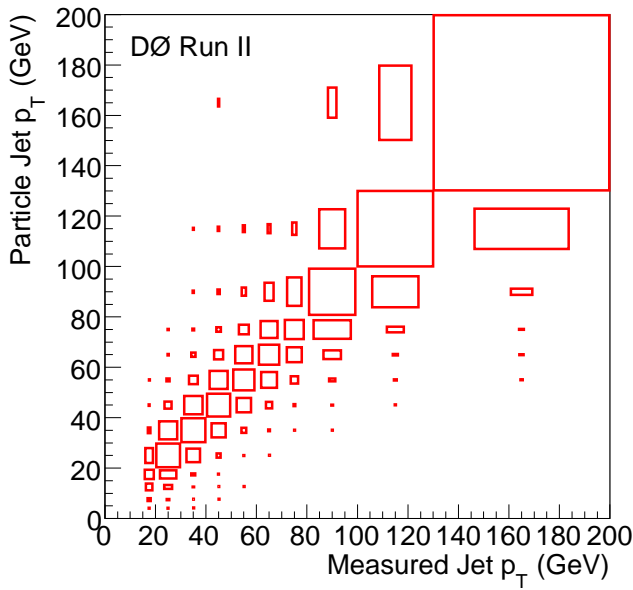}
\caption{\label{fig:matrix}The migration matrix for leading \ptj.
Element $i,j$\ is the probability for a particle jet in \pt\ bin $i$\ to be measured in \pt\ bin $j$, represented by the area of each box. 
Each row sums to unity.  }
\end{figure}

We next describe the process of correcting the \pt\ distribution of the leading (in \pt) jet from the measured level to the particle level; the treatment of \ptz\ is very similar.
The main complexity in the jet \pt\ corrections arises from the experimental \pt\ resolution affecting the relationship between particle jets and the corresponding jets reconstructed in the detector.
First, the finite energy resolution can change the \pt\ ordering of jets between the particle level and detector level.
To account for this we correct the measured \ptJ\ distribution to remove leading measured jets matched to sub-leading particle jets, based on a study of simulated events. 
Here we also remove measured jets arising from additional collisions in the event, modeled by the random bunch crossings from real data overlaid on the simulation.
This combined correction averages ($11.8\pm0.2$)\%, varying from ($33.8\pm0.2$)\% in the range $15<\ptJ<20$~GeV, to ($3.1\pm0.1$)\% for \ptJ$> 50$~GeV.
The second effect of the resolution results in  some jets from a given \ptj\ bin being measured in a different \ptJ\ bin.
This effect is mitigated to a degree by the choice of binning for the measurement: bins are taken to be wider than the detector resolution and to contain a sufficient number of events so that statistical fluctuations do not dominate the final uncertainty on each bin.
Studying the \ptj\ and the corresponding measured \ptJ\ for jets in the full detector simulation allows the remaining effect to be parameterized in a ``migration matrix'' (see Fig. \ref{fig:matrix}), with element $i,j$\ being the probability for a particle jet in \ptj\ bin $i$\ to be measured in \ptJ\ bin $j$. 
The  data distribution is then corrected using a regularized inversion of this matrix~\cite{guru}, with the constraint that the resulting distribution does not have large second derivatives.
Including the reconstructed jets with $15<\ptJ<20$~GeV in the matrix further constrains the effects of lower \ptj\ particle jets fluctuating up in reconstructed \ptJ.
Finally, the distribution is corrected for efficiency and acceptance calculated from simulation, then divided by the bin widths and integrated luminosity to yield the differential cross section.

Uncertainties on the differential cross section are derived empirically through ensemble tests.
A set of 100 ensembles of the same size as the data set are drawn from a \pythia\ sample. 
To reproduce the \ptj\ distribution in data, the \pythia\ \ptj\ spectrum is re-weighted using a function derived from the corrected data and a large (2.5 million events) independent \pythia\ sample. 
Applying this function to the ensembles reproduces the data \ptj\ distribution while retaining realistic statistical fluctuations.
The measured distribution in each ensemble is then corrected in the same way as the data. 
Uncertainties are extracted by taking the fractional difference between the fully corrected distribution and the actual \ptj\ distribution in each ensemble. 
The systematic uncertainty is the mean fractional difference in each \ptj\ bin over all 100 ensembles; the statistical uncertainty is the RMS around the mean.
These statistical uncertainties account for the statistics in each measured bin, and the effects on those statistics of migrations between bins.
The systematic uncertainties are typically below 2\%, and the statistical uncertainty in each bin varies from 2\% at low \ptj\ to 11\%\ at high \ptj.

Further systematic uncertainties are then assessed.
Varying the re-weighting function used in generating the ensembles produces uncertainties at the 3\% level, mostly at low \ptj\ due to the weaker constraints on the particle jet spectrum below the measured region.
Studies of the jet resolution and reconstruction efficiency show small effects, and larger ($\leq3\%$) effects are seen by varying the jet energy scale within uncertainties in the data and simulation.
All other systematic uncertainties studied produced negligible effects.
No strong correlations are observed between the various sources of uncertainty, and the individual contributions are combined in quadrature to obtain the total systematic uncertainty.

The \ptz\ distribution is corrected using the same approach, employing a regularized inversion of the migration matrix with uncertainties derived from ensemble testing; in this case, the muon \pt\ resolution is the source of migration.
Along with the sources of systematic uncertainty considered for the leading jet \pt, the uncertainty on the agreement between the muon resolution in simulation and data is also considered.
Varying the resolution in simulation within uncertainties produces effects below (2--3)\% on the differential cross section.
Varying the jet energy scale produces systematic effects of up to 10\%\ in the region of \ptz$<20$~GeV, which is sensitive to jets close to the reconstructed \ptJ\ cutoff. 

The measurements of \rapz\ and \rapj\ are significantly less challenging, and the method used on these variables is covered briefly.
These distributions do not suffer from significant resolution effects on the rapidity measurement, but still need to be corrected for efficiency and acceptance.
To do this, the ratio of particle level to measured events in each bin is calculated in simulation and applied to the measured data distribution. 
Ensemble testing is then used to measure the uncertainties, with the same sources as the \ptj\ and \ptz\ measurements respectively.
Including the jet systematics covers the correlations between the rapidity and \ptj\ distributions, taking into account changes in the rapidity distributions as events enter or leave the sample due to jet migrations across the \ptJ\ selection of 20~GeV.
As the distributions are symmetric around zero, $|y|$\ is measured in both cases to increase the statistics in each bin.

%% file: results.tex
Integrating over any of the differential cross sections yields the \Z($\rightarrow\mu\mu)$+jet+$X$\ cross section, which we measure to be
$18.7 \pm 0.2 (\rm{stat.}) \pm 0.8(\rm{syst.}) \pm 0.9(\rm{muon}) \pm 1.1 (\rm{lumi.})$~pb, with the following requirements:
 all boson properties are calculated from the muons after QED FSR, and the muons are required to have $|y|<1.7$\ and dimuon mass $65 < M_{\mu\mu} < 115$~GeV; particle jets are reconstructed using the D0 Run~II midpoint algorithm with a splitting/merging fraction of 0.5 and a cone size of $\Delta{\cal R}=0.5$\ on all final state particles except the \Z\ decay products and any FSR photons from the muons, and are required to have $|\rapj| < 2.8$\ and \ptj $> 20$~GeV.
The quoted muon uncertainty covers the muon identification and trigger efficiency determination.
Different definitions of observables complicate comparisons, but this represents a significant reduction in cross section uncertainty from the previously published D0 result using 0.4~fb$^{-1}$\ of data~\cite{d0_zjets}, and is of comparable accuracy to the CDF result using 1.7~fb$^{-1}$~\cite{cdf_zjets}.
For reference, we also measure an inclusive \Z($\rightarrow\mu\mu)$\ cross section requiring only  $65 < M_{\mu\mu} < 115$~GeV, taking the muons after QED FSR with no rapidity requirements, and no jet requirements, to be $233 \pm 1 (\rm{stat.}) \pm 8(\rm{syst.}) \pm 12(\rm{muon}) \pm 14 (\rm{lumi.})$~pb. 
Adding only the requirement that the muons have $|y|<1.7$\ yields a cross section of $118 \pm 0.5 (\rm{stat.}) \pm 4(\rm{syst.}) \pm 6(\rm{muon}) \pm 7 (\rm{lumi.})$~pb.
The muon and luminosity uncertainties are completely correlated between the inclusive \Z\ (with and without the muon $y$\ requirement) and \Z+jet measurements, and all other systematics are uncorrelated.

A number of theoretical predictions are now compared to the measured integrated and differential \Z($\rightarrow\mu\mu)$+jet+$X$\ cross sections.
NLO pQCD calculations are obtained using \mcfm\ together with the NLO CTEQ6.6M parton distribution functions of the proton (PDF)~\cite{cteq66}. 
The associated PDF error sets are used to assess uncertainties, which are about 3\%.
Renormalization and factorization scales are set to the sum in quadrature of the mass and \pt\ of the $Z$~boson, and uncertainties are derived by varying  both scales down or up together by a factor of two, which changes the prediction by $\pm 7$\%.
Non-perturbative corrections for hadronization and the underlying event are derived from \pythia\ v6.418 \Z\ + jet production, with the leading order (LO) PDF CTEQ5L~\cite{cteq5L} and the underlying event tune DW~\cite{pythia_tune}.
These are derived by comparing the full prediction (taken from the final state particles, including the underlying event) to the purely perturbative part (calculated from partons taken after the parton shower, with no underlying event). 
Corrections for QED FSR from the muons are derived from the same \pythia\ sample, by comparing the prediction calculated using the muons after FSR to those using the muons before FSR. 
These FSR corrections are around 2\% caused mainly by events migrating out of the mass window, with little dependence on any variable considered except at low \ptz\ and high \rapz.
The non-perturbative and FSR corrections are given in Tables \ref{tab:mcfm_jet_pt}, \ref{tab:mcfm_jet_rap}, \ref{tab:mcfm_z_pt}, and \ref{tab:mcfm_z_rap}. 
The prediction for the \Z+jet+$X$\ cross section from NLO pQCD with our stated acceptance cuts and after corrections (hereafter referred to as the NLO pQCD prediction), is $17.3  \pm 1.2 (\rm{scale}) \pm 0.5 (\rm{PDF})$~pb.
This is 5\% below the measured value, and within uncertainties. 
For reference, \mcfm\ is also used to calculate the LO pQCD prediction.
After acceptance cuts and corrections this is $12.8 ^{+2.1}_{-1.7} (\rm{scale}) \pm 0.3 (\rm{PDF})$~pb.


\begin{table}
\caption{\label{tab:mcfm_jet_pt}Non-perturbative and FSR correction factors applied to the \mcfm~prediction for leading \ptj\ in \Z\ + jet + $X$\ events.}
\begin{ruledtabular}
\begin{tabular}{D{,}{\,-\,}{-1}cc}
\multicolumn{1}{c} \mbox{\ptj~~~} & Non-pert. & FSR \\
\multicolumn{1}{c} \mbox{(GeV)~} & Corr.   & Corr.  \\
    \hline
20,30 & 1.041 & 0.977 \\
30,40 & 1.017 & 0.977 \\
40,50 & 1.001 & 0.977 \\
50,60 & 0.995 & 0.977 \\
60,70 & 0.991 & 0.977 \\
70,80 & 0.989 & 0.978 \\
80,100 & 0.988 & 0.978 \\
100,130 & 0.986 & 0.978 \\
130,200 & 0.984 & 0.978 \\
\end{tabular}
\end{ruledtabular}
\end{table}

\begin{table}
\caption{\label{tab:mcfm_jet_rap}Non-perturbative and FSR correction factors applied to the \mcfm~prediction for leading $|\rapj|$\ in \Z\ + jet + $X$\ events.
}
\begin{ruledtabular}
\begin{tabular}{D{,}{\,-\,}{-1}cc}
\multicolumn{1}{c} \mbox{$|\rapj|$~~} & Non-pert. & FSR \\
                         & Corr.   & Corr.  \\
    \hline
0.0,0.4 & 1.026 & 0.977 \\
0.4,0.8 & 1.025 & 0.977 \\
0.8,1.2 & 1.022 &  0.977 \\
1.2,1.6 & 1.018 & 0.977 \\
1.6,2.0 & 1.001 & 0.977 \\
2.0,2.4 & 1.009 & 0.977 \\
2.4,2.8 & 0.983 & 0.977 \\
\end{tabular}
\end{ruledtabular}
\end{table}

\begin{table}
\caption{\label{tab:mcfm_z_pt}Non-perturbative and FSR correction factors applied to the \mcfm~prediction for \ptz\ in \Z\ + jet + $X$\ events.}
\begin{ruledtabular}
\begin{tabular}{D{,}{\,-\,}{-1}cc}
\multicolumn{1}{c} \mbox{\ptz~~~} & Non-pert. & FSR \\
\multicolumn{1}{c} \mbox{(GeV)~} & Corr.   & Corr.  \\
    \hline
0,10 &   1.099 & 1.268 \\
10,18 & 1.284 & 1.041 \\
18,26 & 1.021 & 0.977 \\
26,35 & 0.995 & 0.972 \\
35,45 & 0.997 & 0.973 \\
45,60 & 0.999 & 0.967 \\
60,80 & 1.000 & 0.959 \\
80,120 & 1.000 & 0.951 \\
120,200 & 1.000 & 0.944 \\
\end{tabular}
\end{ruledtabular}
\end{table}

\begin{table}
\caption{\label{tab:mcfm_z_rap}Non-perturbative and FSR correction factors applied to the \mcfm~prediction for $|\rapz|$\ in \Z\ + jet + $X$\ events.}
\begin{ruledtabular}
\begin{tabular}{D{,}{\,-\,}{-1}cc}
\multicolumn{1}{c} \mbox{$|\rapz|$~~} & Non-pert. & FSR \\
 & Corr.   & Corr.  \\
    \hline
0.0,0.2 & 1.024 & 0.979 \\
0.2,0.4 & 1.024 & 0.978 \\
0.4,0.6 & 1.024 & 0.978 \\
0.6,0.8 & 1.021 & 0.978 \\
0.8,1.0 & 1.017 & 0.977 \\
1.0,1.2 & 1.013 & 0.975 \\
1.2,1.4 & 1.010 & 0.971 \\
1.4,1.6 & 1.001 & 0.965 \\
1.6,1.8 & 1.000 & 0.928 \\
\end{tabular}
\end{ruledtabular}
\end{table}


Comparisons are also made using three event generators:
i) a sample generated with \alpgen\ v2.13 with up to three partons in the matrix element calculation, and the factorization and renormalization scales squared set to the sum in quadrature of the mass and \pt\ of the $Z$~boson, and \pythia\ with tune QW~\cite{pythia_tune} used for the parton showering;
ii) a sample generated with \sherpa\ v1.1.1, again with up to three partons in the matrix element calculation, and the default parton showering algorithm ({\sc apacic}) and underlying event model ({\sc amisic});
iii) an inclusive \Z\ sample generated with \pythia\ v6.418, using tune QW. 
In order to use consistent PDF with all three, the NLO CTEQ6.1M~\cite{cteq61} is chosen.
For both \alpgen\ and \sherpa, jets from the matrix element calculation are required to  have $p_{T}>15$~GeV, and a separation $\Delta R>0.4$, to ensure full coverage of the measured phase space.
For all three event generators, the boson kinematics are calculated from the muons after QED FSR, for consistency with the measured observables.
After applying our stated particle level \Z + jet selection, the predicted cross sections are 11.6~pb (\alpgen), 15.0~pb (\sherpa), and 12.1~pb (\pythia).


\begin{figure}[!h]\center
\includegraphics[width=85mm]{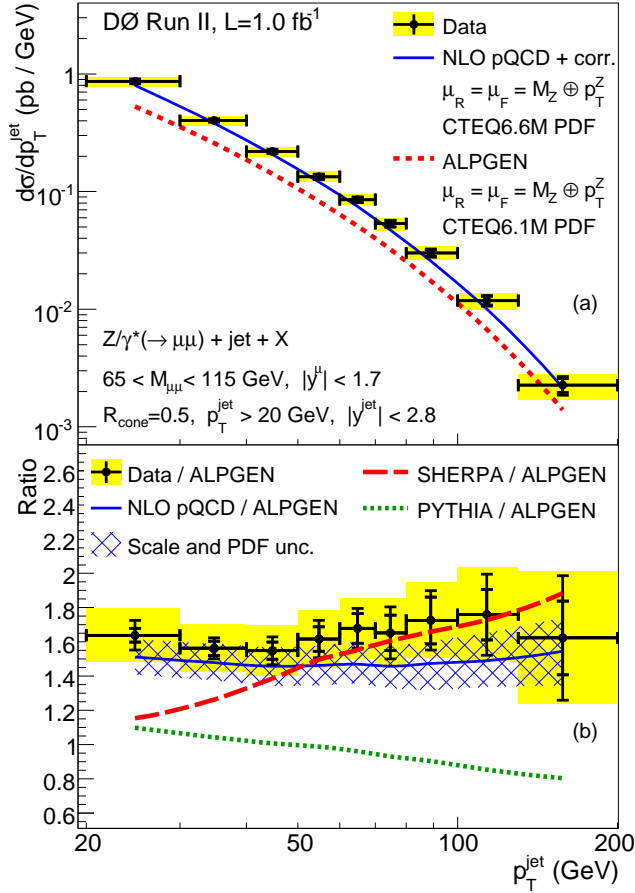}
\caption{\label{fig:jet_pt_result}(a) The measured cross section in bins of leading \ptj\ for \Z\ + jet + $X$\ events. Predictions from NLO pQCD and \alpgen\ are compared to the data. 
(b) The ratio of data and predictions from NLO pQCD + corrections, \sherpa, and \pythia\ to the prediction from \alpgen.
}
\end{figure}

\begin{figure}[!h]\center
\includegraphics[width=85mm]{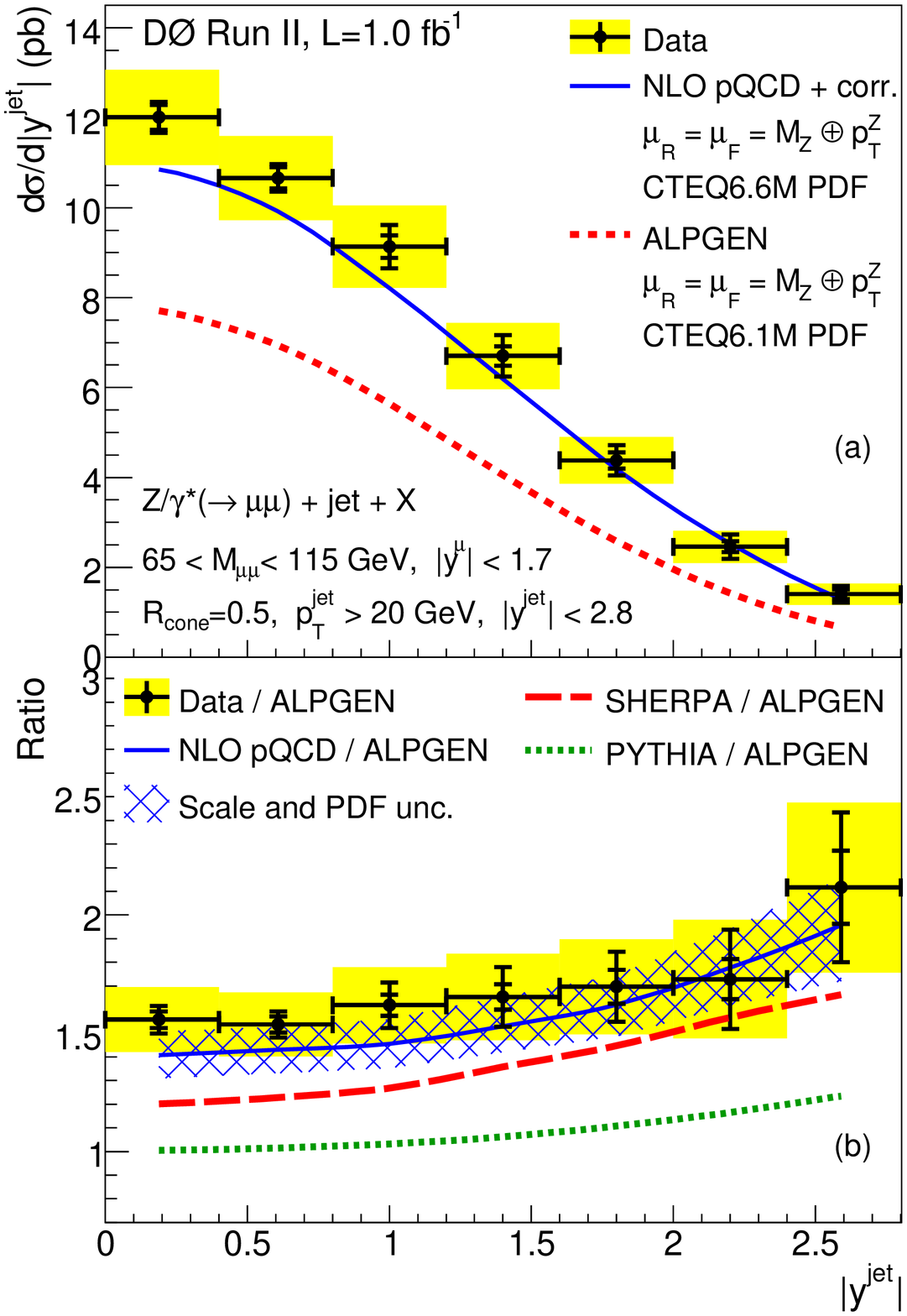}
\caption{\label{fig:jet_rap_result}(a) The measured cross section in bins of leading $|\rapj|$\ for \Z\ + jet + $X$\ events. Predictions from NLO pQCD and \alpgen\ are compared to the data.  
(b) The ratio of data and predictions from NLO pQCD + corrections, \sherpa, and \pythia\ to the prediction from \alpgen.}
\end{figure}

\begin{figure}[!h]\center
\includegraphics[width=85mm]{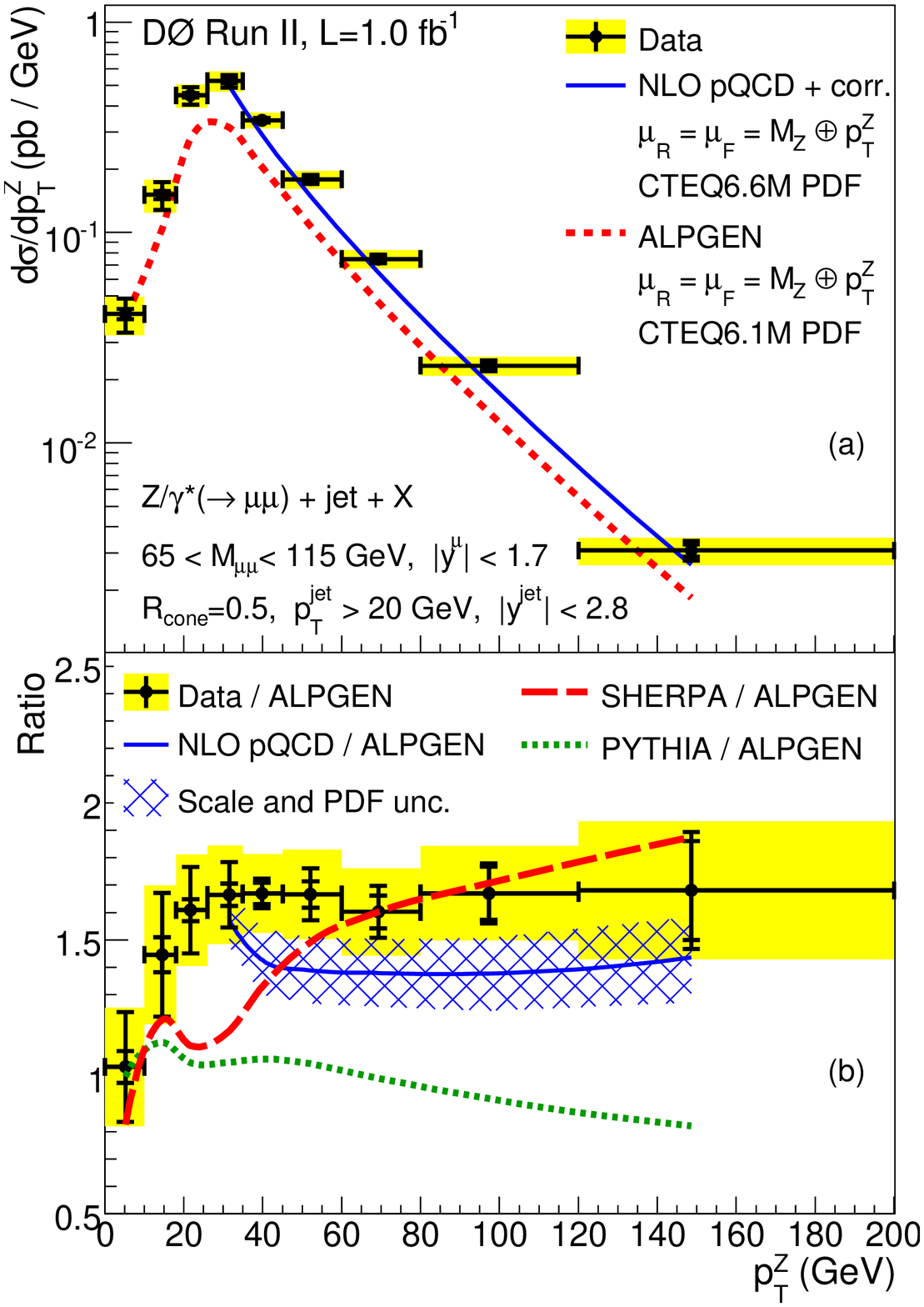}
\caption{\label{fig:z_pt_result}(a) The measured cross section in bins of \ptz\ for \Z\ + jet + $X$\ events. Predictions from NLO pQCD and \alpgen\ are compared to the data. 
(b) The ratio of data and predictions from NLO pQCD + corrections, \sherpa, and \pythia\ to the prediction from \alpgen.}
\end{figure}

\begin{figure}[!h]\center
\includegraphics[width=85mm]{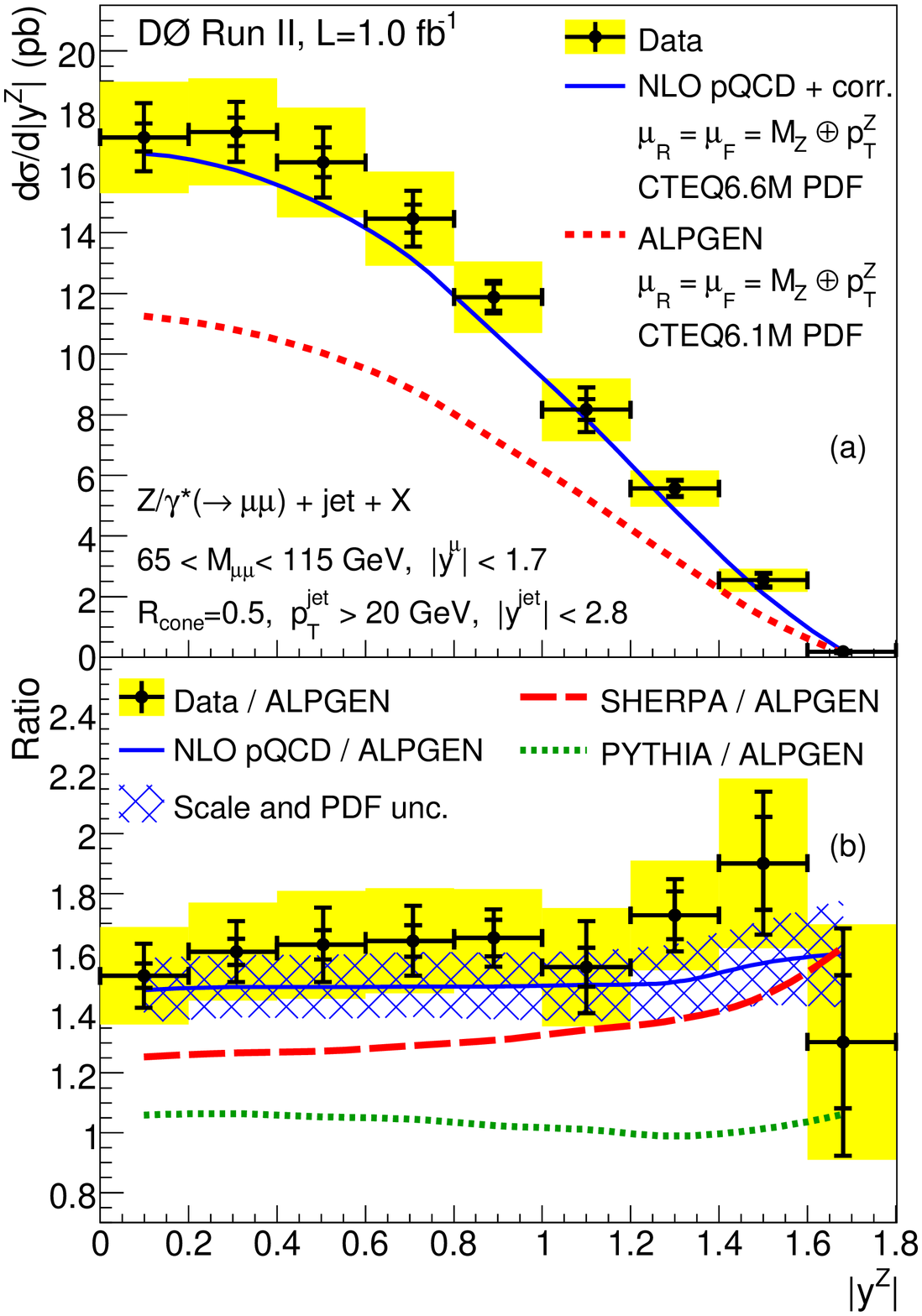}
\caption{\label{fig:z_rap_result}(a) The measured cross section in bins of $|\rapz|$\ for \Z\ + jet + $X$\ events. Predictions from NLO pQCD and \alpgen\ are compared to the data. 
(b) The ratio of data and predictions from NLO pQCD + corrections, \sherpa, and \pythia\ to the prediction from \alpgen.}
\end{figure}

The differential cross sections are shown binned in leading \ptj\ (Fig.\ \ref{fig:jet_pt_result}), leading jet \rap\  (Fig.\ \ref{fig:jet_rap_result}),  \ptz\ in events with at least one jet (Fig.\ \ref{fig:z_pt_result}), and \rapz\ in events with at least one jet (Fig.\ \ref{fig:z_rap_result}).
Data points in each bin are placed where the differential cross section in simulation is equal to the bin average~\cite{bins}.
The data are shown with statistical uncertainties (inner error bar) and sum in quadrature of statistical and systematic uncertainties (outer error bar), excluding the uncertainties on the measured integrated luminosity and the muon identification and trigger efficiencies. 
These final two uncertainties are completely correlated between bins and with the muon and luminosity uncertainties on the measured inclusive \Z\ cross section; however, they are included to form the total uncertainty, shown as the shaded region. 
For clarity, only the predictions of NLO pQCD  and \alpgen\ are shown in part (a) of each figure, though the prediction from NLO pQCD is not shown at low \ptz\ (Fig.~\ref{fig:z_pt_result}) where non-perturbative processes dominate over the NLO contribution.
The data results are also provided in Tables \ref{tab:shape_unc_jetpt}, \ref{tab:shape_unc_jety}, \ref{tab:shape_unc_zpt}, and \ref{tab:shape_unc_zy}.
In part (b) of each figure, the distributions from data, NLO pQCD, \sherpa\ and \pythia\ are shown divided by the prediction from \alpgen. 
The NLO pQCD prediction is shown with the scale and PDF uncertainties combined in quadrature as a hatched region; the scale uncertainty is approximately a factor of two larger than the PDF uncertainty across all distributions.


\begin{table}
\caption{\label{tab:shape_unc_jetpt}The measured cross section in bins of leading \ptj\ for \Z\ + jet + $X$\ events.
Uncertainties are split into statistical and systematic; these are combined with an additional constant 8.0\% normalization uncertainty from the luminosity, trigger, and muon identification to form the total uncertainty.
}
\begin{ruledtabular}
\begin{tabular}{D{,}{\,-\,}{-1}D{,}{\,.\,}{-1}D{,}{\,.\,}{-1}D{,}{\,.\,}{-1}D{,}{\,.\,}{-1}D{,}{\,.\,}{-1}}
\multicolumn{1}{c} \mbox{\ptj~~ } & 
\multicolumn{1}{c} \mbox{Bin ctr.} & 
\multicolumn{1}{c} \mbox{$\rm{d}\sigma/\rm{d}\ptj$} & 
\multicolumn{1}{c} \mbox{$\delta\sigma_{\rm{stat.}}$}  &
\multicolumn{1}{c} \mbox{$\delta\sigma_{\rm{syst.}}$} &
\multicolumn{1}{c} \mbox{$\delta\sigma_{\rm{total}}$} \\
\multicolumn{1}{c} \mbox{(GeV)~ } &  
\multicolumn{1}{c} \mbox{(GeV)~} &  
\multicolumn{1}{c} \mbox{(pb/GeV)}   & 
\multicolumn{1}{c} \mbox{(\%)~}   &  
\multicolumn{1}{c} \mbox{(\%)~} &  
\multicolumn{1}{c} \mbox{(\%)~}  \\
    \hline
20,30 & 24,7 & 0,867 & 2,4 & 4,7 & 9,6 \\
30,40 & 34,8 & 0,402 & 2,7 & 2,9 & 8,9 \\
40,50 & 44,8 & 0,219 & 3,2 & 4,2 & 9,5 \\
50,60 & 54,8 & 0,134 & 4,4 & 5,1 & 10,5 \\
60,70 & 64,8 & 0,0854 & 5,1 & 4,6 & 10,6 \\
70,80 & 74,7 & 0,0535 & 6,3 & 6,7 & 12,2 \\
80,100 & 89,0 & 0,0301 & 7,9 & 6,2 & 12,9 \\
100,130 & 113,5 & 0,0118 & 8,4 & 10,6 & 15,7 \\
130,200 & 157,7 & 0,00226 & 13,3 & 18,1 & 23,8 \\
\end{tabular}
\end{ruledtabular}
\end{table}

\begin{table}
\caption{\label{tab:shape_unc_jety}The measured cross section in bins of leading $|\rapj|$\ for \Z\ + jet + $X$\ events.
Uncertainties are split into statistical and systematic; these are combined with  an additional constant 8.0\% normalization uncertainty from the luminosity, trigger, and muon identification to form the total uncertainty.
}
\begin{ruledtabular}
\begin{tabular}{D{,}{\,-\,}{-1}D{,}{\,.\,}{-1}D{,}{\,.\,}{-1}D{,}{\,.\,}{-1}D{,}{\,.\,}{-1}D{,}{\,.\,}{-1}}
\multicolumn{1}{c} \mbox{$|\rapj|$~~} & 
\multicolumn{1}{c} \mbox{Bin~~} & 
\multicolumn{1}{c} \mbox{$\rm{d}\sigma/\rm{d}|\rapj|$} & 
\multicolumn{1}{c} \mbox{$\delta\sigma_{\rm{stat.}}$}  &
\multicolumn{1}{c} \mbox{$\delta\sigma_{\rm{syst.}}$} &
\multicolumn{1}{c} \mbox{$\delta\sigma_{\rm{total}}$} \\
 &  
\multicolumn{1}{c} \mbox{center}  & 
\multicolumn{1}{c} \mbox{(pb)~~}   & 
\multicolumn{1}{c} \mbox{(\%)~}   &  
\multicolumn{1}{c} \mbox{(\%)~} &  
\multicolumn{1}{c} \mbox{(\%)~}  \\
    \hline
0.0,0.4 & 0,189 & 12,01 & 2,3 & 2,9 & 8,8 \\
0.4,0.8 & 0,609 & 10,66 & 2,2 & 2,9 & 8,8 \\
0.8,1.2 & 1,00 & 9,13 & 2,8 & 5,3 & 10,0 \\
1.2,1.6 & 1,40 & 6,70 & 3,2 & 6,9 & 11,0 \\
1.6,2.0 & 1,80 & 4,38 & 4,2 & 7,6 & 11,8 \\
2.0,2.4 & 2,20 & 2,46 & 4,9 & 11,1 & 14,5 \\
2.4,2.8 & 2,59 & 1,40 & 7,3 & 13,1 & 17,0 \\
\end{tabular}
\end{ruledtabular}
\end{table}

\begin{table}
\caption{\label{tab:shape_unc_zpt}The measured cross section in bins of \ptz\ for \Z\ + jet + $X$\ events.
Uncertainties are split into statistical and systematic; these are combined with an additional constant 8.0\% normalization uncertainty from the luminosity, trigger, and muon identification to form the total uncertainty.
}
\begin{ruledtabular}
\begin{tabular}{D{,}{\,-\,}{-1}D{,}{\,.\,}{-1}D{,}{\,.\,}{-1}D{,}{\,.\,}{-1}D{,}{\,.\,}{-1}D{,}{\,.\,}{-1}}
\multicolumn{1}{c} \mbox{\ptz~~~} & 
\multicolumn{1}{c} \mbox{~~Bin ctr.} & 
\multicolumn{1}{c} \mbox{$\rm{d}\sigma/\rm{d}\ptz$~} & 
\multicolumn{1}{c} \mbox{$\delta\sigma_{\rm{stat.}}$}  &
\multicolumn{1}{c} \mbox{$\delta\sigma_{\rm{syst.}}$} &
\multicolumn{1}{c} \mbox{$\delta\sigma_{\rm{total}}$} \\
\multicolumn{1}{c} \mbox{(GeV)~} &  
\multicolumn{1}{c} \mbox{(GeV)~} &  
\multicolumn{1}{c} \mbox{(pb/GeV)}   & 
\multicolumn{1}{c} \mbox{(\%)~}   &  
\multicolumn{1}{c} \mbox{(\%)~} &  
\multicolumn{1}{c} \mbox{(\%)~}  \\
    \hline
0,10 & 5,2 & 0,0410 & 5,6 & 18,5 & 20,1 \\
10,18 & 14,5 & 0,151 & 4,4 & 15,0 & 17,0 \\
18,26 & 21,7 & 0,448 & 2,5 & 9,5 & 12,6 \\
26,35 & 31,5 & 0,525 & 2,5 & 6,8 & 10,8 \\
35,45 & 39,8 & 0,342 & 2,3 & 2,2 & 8,6 \\
45,60 & 52,1 & 0,179 & 2,9 & 4,9 & 9,8 \\
60,80 & 69,3 & 0,0748 & 3,7 & 4,6 & 9,9 \\
80,120 & 97,3 & 0,0233 & 5,8 & 3,0 & 10,3 \\
120,200 & 148,6 & 0,00309 & 10,8 & 6,7 & 15,0 \\

\end{tabular}
\end{ruledtabular}
\end{table}

\begin{table}
\caption{\label{tab:shape_unc_zy}The measured cross section in bins of $|\rapz|$\ for \Z\ + jet + $X$\ events.
Uncertainties are split into statistical and systematic; these are combined with an additional constant 8.0\% normalization uncertainty from the luminosity, trigger, and muon identification to form the total uncertainty.
}
\begin{ruledtabular}
\begin{tabular}{D{,}{\,-\,}{-1}D{,}{\,.\,}{-1}D{,}{\,.\,}{-1}D{,}{\,.\,}{-1}D{,}{\,.\,}{-1}D{,}{\,.\,}{-1}}
\multicolumn{1}{c} \mbox{$|\rapz|$~~} & 
\multicolumn{1}{c} \mbox{Bin~~} & 
\multicolumn{1}{c} \mbox{~~$\rm{d}\sigma/\rm{d}|\rapz|$} & 
\multicolumn{1}{c} \mbox{$\delta\sigma_{\rm{stat.}}$}  &
\multicolumn{1}{c} \mbox{$\delta\sigma_{\rm{syst.}}$} &
\multicolumn{1}{c} \mbox{$\delta\sigma_{\rm{total}}$} \\
 & 
\multicolumn{1}{c} \mbox{center} &  
\multicolumn{1}{c} \mbox{(pb)~~}   & 
\multicolumn{1}{c} \mbox{(\%)~}   &  
\multicolumn{1}{c} \mbox{(\%)~} &  
\multicolumn{1}{c} \mbox{(\%)~}  \\
    \hline
0.0,0.2 & 0,099 & 17,15 & 2,7 & 6,5 & 10,6 \\
0.2,0.4 & 0,308 & 17,33 & 2,7 & 5,7 & 10,2 \\
0.4,0.6 & 0,504 & 16,32 & 3,0 & 7,1 & 11,1 \\
0.6,0.8 & 0,708 & 14,47 & 3,2 & 6,4 & 10,7 \\
0.8,1.0 & 0,890 & 11,88 & 3,7 & 4,5 & 9,9 \\
1.0,1.2 & 1,10  & 8,17 & 4,2 & 9,0 & 12,7 \\
1.2,1.4 & 1,30  & 5,57 & 4,6 & 5,2 & 10,6 \\
1.4,1.6 & 1,50  & 2,54 & 8,2 & 9,6 & 14,9 \\
1.6,1.8 & 1,68  & 0,17 & 17,0 & 23,6 & 30,2 \\
\end{tabular}
\end{ruledtabular}
\end{table}

%% file: summary.tex
In summary, we have measured  differential cross sections for \Z+jet+$X$\ production with $0.97\pm0.06$~fb$^{-1}$\ of integrated luminosity recorded by the D0 experiment in \pp\ collisions at \roots.
We presented the first results binned in \ptz\ and \rapz; as well as new results binned in leading jet \pt\ and \rap, extending the measured \ptj\ and \rapj\ ranges substantially.
The total \Z+jet+$X$\ cross section measured in data is 5\%\ above the prediction from NLO pQCD + corrections, which is within the total uncertainties. 
This is comparable with the trend observed in previous measurements~\cite{cdf_zjets}, although direct comparisons are complicated by different definitions of the final observables.
The shapes of the differential distributions are generally well described by NLO pQCD, though the distribution at lower \ptz\ (below the \ptj\ cutoff of 20~GeV) is dominated by non-perturbative processes.
The total cross sections predicted by the \alpgen\ and \pythia\ event generators are significantly below the measured value and are more consistent with the LO pQCD predictions.
The prediction from \sherpa\ lies between the LO and NLO pQCD prediction.
The shapes of the data distributions are generally well described by \alpgen, except at low \ptz. 
There is also indication that the jet rapidity distribution is narrower in \alpgen\ than in data, NLO pQCD, \sherpa\ and \pythia.
Comparisons to the other event generators show that \sherpa\ has a slope in \ptj\  and \ptz\ relative to the data, with more events at high \pt\ compared to low \pt; \pythia\ shows the opposite behavior.
This measurement tests the current best predictions for heavy boson + jet production at hadron colliders.
As the data are fully corrected for instrumental effects, they can be directly used in testing and improving the existing event generators, or any future calculations and models.

%% file: acknowledgement_paragraph_r2.tex
%
We thank the staffs at Fermilab and collaborating institutions, 
and acknowledge support from the 
DOE and NSF (USA);
CEA and CNRS/IN2P3 (France);
FASI, Rosatom and RFBR (Russia);
CNPq, FAPERJ, FAPESP and FUNDUNESP (Brazil);
DAE and DST (India);
Colciencias (Colombia);
CONACyT (Mexico);
KRF and KOSEF (Korea);
CONICET and UBACyT (Argentina);
FOM (The Netherlands);
STFC (United Kingdom);
MSMT and GACR (Czech Republic);
CRC Program, CFI, NSERC and WestGrid Project (Canada);
BMBF and DFG (Germany);
SFI (Ireland);
The Swedish Research Council (Sweden);
CAS and CNSF (China);
and the
Alexander von Humboldt Foundation (Germany).

%% file: publication.bbl
\begin{thebibliography}{99}
\bibitem{rapidity}Rapidity is defined as $y = \ln\frac{E-p_z}{E+p_z}$, where $E$\ is the energy, and $p_z$\ the component of momentum parallel to the proton beam direction.
\bibitem{theta} $\eta = -\ln [\tan(\theta/2)]$, where $\theta$\ is the polar angle, defined with respect to the proton beam axis.
\end{thebibliography}
